\documentclass[a4paper,english]{lipics-v2019}
\usepackage{nth}
\usepackage{microtype}
\usepackage{breqn}
\usepackage{ifthen}
\usepackage{placeins}
\usepackage{complexity}
\usepackage{nicefrac}
\usepackage{todonotes}
\usepackage{siunitx}
\usepackage{amssymb}
\usepackage{color}
\usepackage{gensymb}

\hideLIPIcs
\nolinenumbers

\newcommand{\old}[1]{{}}
\newcommand{\change}[1]{{\color{red}#1}}

\title{Minimum Coverage by Convex Polygons:\\ The CG:SHOP Challenge 2023}
\titlerunning{Minimum Coverage by Convex Polygons: CG Challenge 2023}

\author{Sándor P.~Fekete}{Department of Computer Science, TU Braunschweig, Germany}{s.fekete@tu-bs.de}{https://orcid.org/0000-0002-9062-4241}{}
\author{Phillip Keldenich}{Department of Computer Science, TU Braunschweig, Germany}{p.keldenich@tu-bs.de}{https://orcid.org/0000-0002-6677-5090}{}
\author{Dominik Krupke}{Department of Computer Science, TU Braunschweig, Germany}{d.krupke@tu-bs.de}{https://orcid.org/0000-0003-1573-3496}{}
\author{Stefan Schirra}{Department for Simulation and Graphics, OvGU Magdeburg, Germany}{stschirr@isg.cs.uni-magdeburg.de}{https://orcid.org/0009-0006-5928-1494}{}
\authorrunning{S.~P.~Fekete, P.~Keldenich, D.~Krupke, S.~Schirra}
\Copyright{S.~P.~Fekete, P.~Keldenich, D.~Krupke, S.~Schirra}



\supplement{}

\funding{Work at TU Braunschweig has been partially supported by the German Research Foundation (DFG), 
project ``Computational Geometry: Solving Hard Optimization Problems'' (CG:SHOP), grant FE407/21-1.} 

\acknowledgements{We thank the members of the CG:SHOP Challenge Advisory Board for their valuable input: 
William J.~Cook, Andreas Fabri, Dan Halperin, Michael Kerber, Philipp Kindermann, Joe Mitchell, Kevin Verbeek.
Further thanks go to Efi Fogel for consulting us regarding a problem with CGAL, and to Rouven Kniep for implementing a simple greedy approach that served us as a baseline for the evaluation.}

\EventAcronym{CG:SHOP 2023}

\begin{document}
\maketitle
\begin{abstract}
We give an overview of the 2023 Computational Geometry Challenge
targeting the problem {\sc Minimum Coverage by Convex Polygons},
which consists of covering a given polygonal region (possibly with holes)
by a minimum number of convex subsets, a problem with a long-standing
tradition in Computational Geometry.
\end{abstract}

\section{Introduction}
The ``CG:SHOP Challenge'' (Computational Geometry: Solving Hard
Optimization Problems) originated as a workshop at the 2019
Computational Geometry Week (CG Week) in Portland, Oregon in June,
2019.  The goal was to conduct a computational challenge competition
that focused attention on a specific hard geometric optimization
problem, encouraging researchers to devise and implement solution
methods that could be compared scientifically based on how well they
performed on a database of carefully selected and varied instances.
While much of computational
geometry research is theoretical, often seeking provable approximation
algorithms for \NP-hard optimization problems,
the goal of the Challenge was to set the metric of success based on
computational results on a specific set of benchmark geometric
instances. The 2019 Challenge focused on the problem of computing
simple polygons of minimum and maximum area for given sets of vertices in the
plane. This Challenge generated a strong response from many research
groups, from both the computational geometry and the combinatorial
optimization communities, and resulted in a lively exchange of
solution ideas.

For CG Weeks 2020, 2021, and 2022 the Challenge problems were {\sc Minimum Convex Partition},
{\sc Coordinated Motion Planning}, and {\sc Minimum Partition into Plane Subgraphs}, 
respectively. The CG:SHOP Challenge became an event within
the CG Week program, with top performing solutions reported in the
Symposium on Computational Geometry (SoCG) proceedings. The schedule for the
Challenge was advanced earlier, to give an opportunity for more
participation, particularly among students, e.g., as part of course
projects. 

The fifth edition of the Challenge in 2023 continued
this format, leading to contributions in the SoCG proceedings.
A total of 22 teams registered, with 18 submitting at least one valid solution.

\section{The Challenge: Minimum Coverage by Convex Polygons}

A suitable contest problem has a number of desirable properties.

\begin{itemize}
\item The problem is of geometric nature.
\item The problem is of general scientific interest and has received previous attention.
\item Optimization problems tend to be more suitable than feasibility problems; in principle, 
  feasibility problems are also possible, but they need to be suitable for sufficiently
  fine-grained scoring to produce an interesting contest.
\item Computing optimal solutions is difficult for instances of reasonable size.
\item This difficulty is of a fundamental algorithmic nature, and not only due to
 issues of encoding or access to sophisticated software or hardware.
\item Verifying feasibility of provided solutions is relatively easy.
\end{itemize}

In this fifth year, a call for suitable problems was communicated in June 2022.
In response, a total of six interesting problems were proposed for the 2023 Challenge.
These were evaluated with respect to difficulty, distinctiveness from previous years,
and existing literature and related work. In the end, the Advisory Board
selected the chosen problem. Special thanks go to Dan Halperin (Tel Aviv University)
who suggested this problem, motivated by applications from the field of Robotics~\cite{AmiceDWZT22}.

\subsection{The Problem}
The specific problem that formed the basis of the 2023 CG Challenge was the following;
see Figure~\ref{fig:example} for a simple example.
\change{
\begin{figure}[h]
	\centering
	\includegraphics{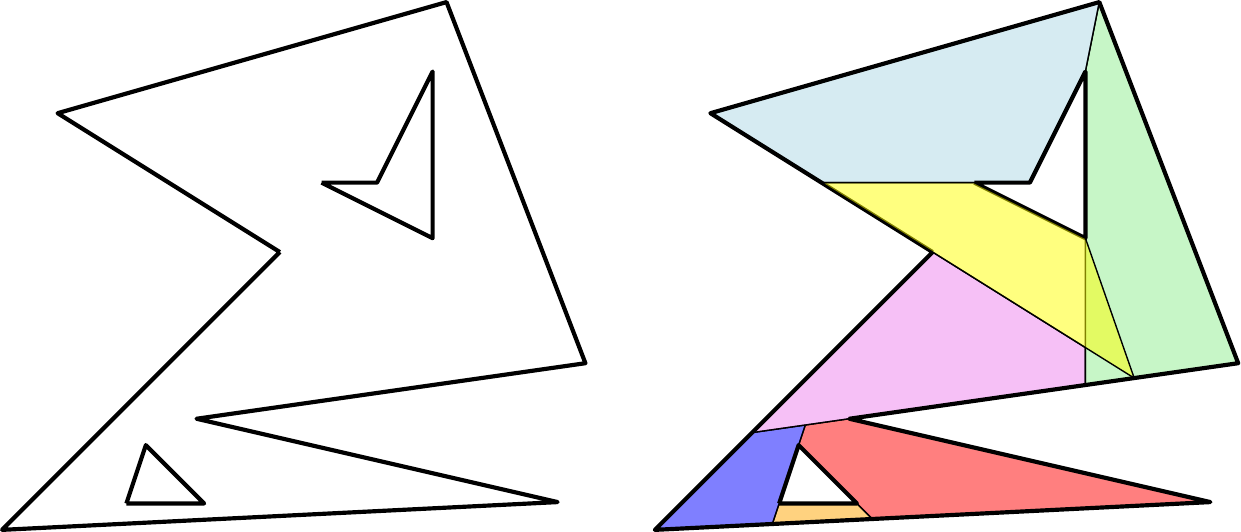}
	\caption{A possible instance, given by a (non-simple) polygonal region in the plane (left),
			 and a feasible cover by convex sets (right).}
	\label{fig:example}
\end{figure}
}

\medskip
\noindent \textbf{Problem:} {\sc Minimum Coverage by Convex Polygons} 

\noindent \textbf{Given:} Given a geometric region, $P$, in the plane, which 
may be a simple polygon or a polygon with holes. 

\noindent \textbf{Goal:} The task is to cover
$P$ with a collection, $C_1,\ldots,C_k$ of convex polygons, each contained within
$P$, such that the number $k$ of convex polygons in the cover is minimized.

\begin{figure}
	\centering
	\includegraphics[width=.8\textwidth]{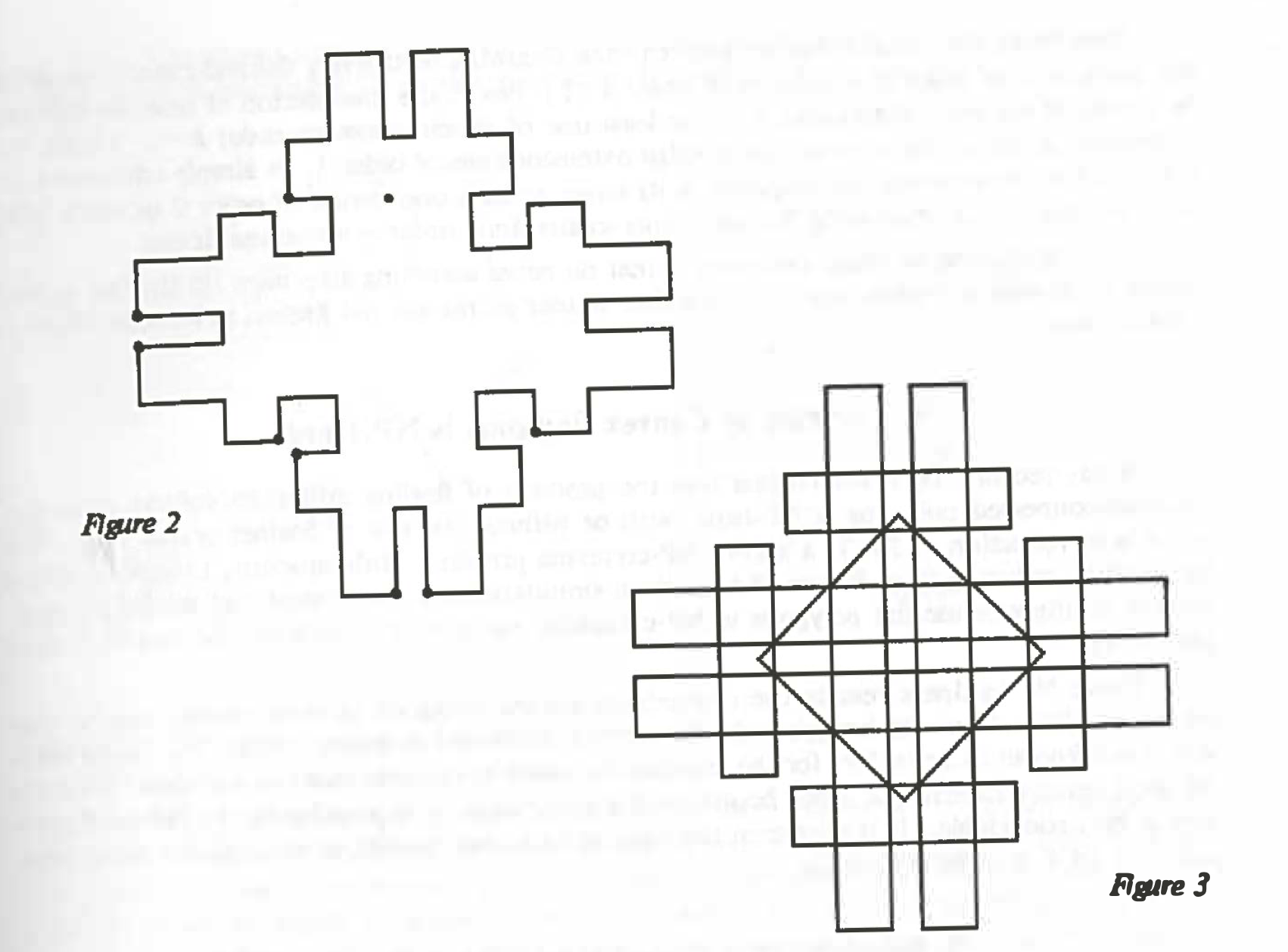}
	\includegraphics[width=.8\textwidth]{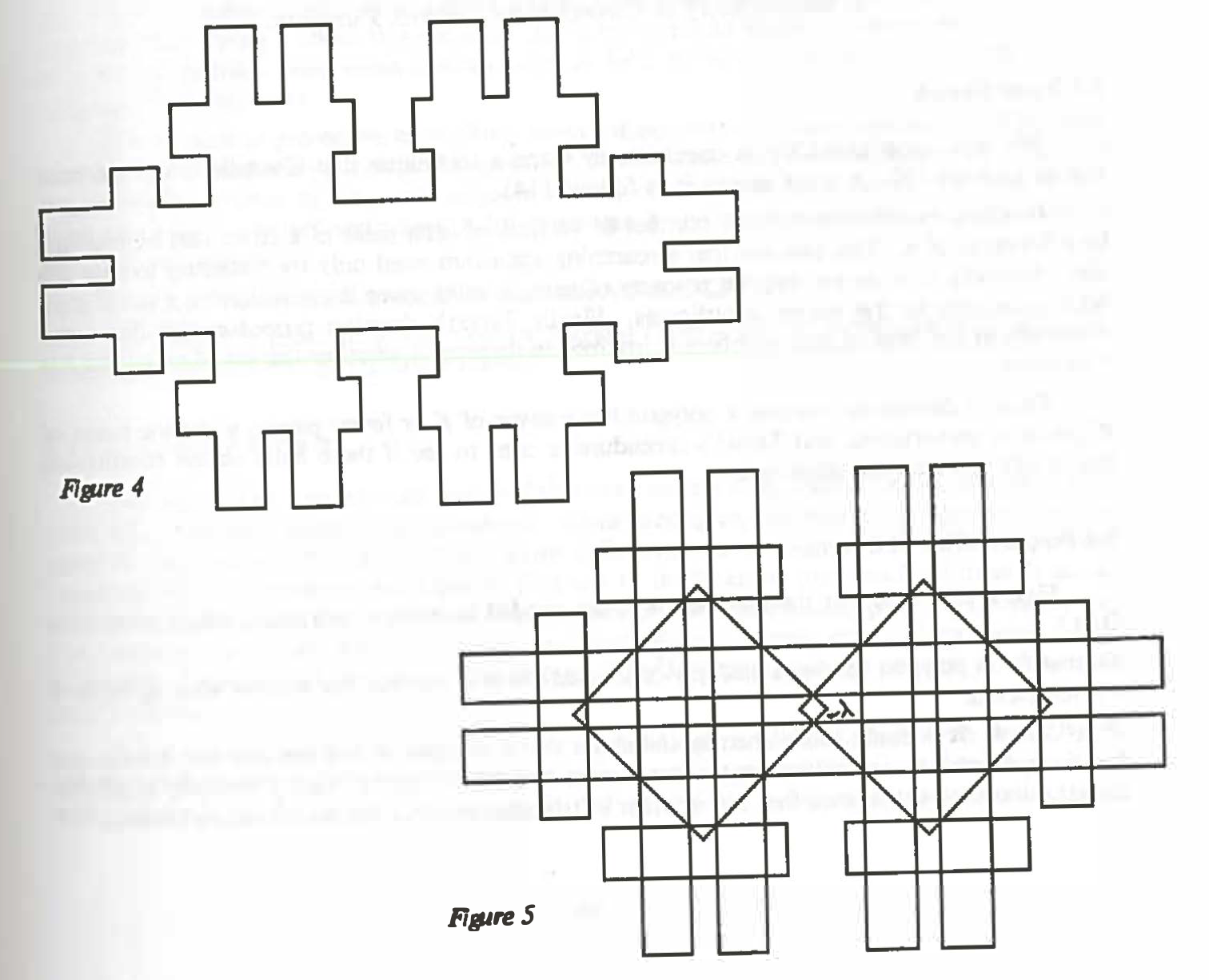}
	\caption{The SoCG logo and minimum convex covering. A simple polygonal region (top left, ``Figure 2'') and a minimum convex cover consisting of
	nine convex polygons, one of which is a diamond (i.e., a rotated square) that does not have a vertex on an edge of the region (top right, ``Figure 3'').
	Combining two such polygons into one (bottom left, ``Figure 4'') results in a region for which an optimal solution requires
	Steiner points that do not even lie on edge extensions (bottom right, ``Figure 5''). All images from O'Rourke~\cite{ORourkeMCC82}.}
	\label{fig:logo}
\end{figure}

Variants of this problem have a long history in Computational Geometry; in fact,
a duplicated logo of the annual SoCG conference (shown in Figure~\ref{fig:logo}) illustrates
that even for a simple polygonal region with axis-parallel edges, 
a minimum convex cover may need to employ vertices
that do not lie on the arrangement of extensions of the
edges of the polygon; see the paper~\cite{ORourkeMCC82} by O'Rourke, the program 
chair of the first conference in 1985, and \url{https://www.computational-geometry.org/logo.html}.

\subsection{Related Work}

Even in the early years of computational geometry,
convex covering gained much attention~\cite{KEIL1985197}.
The same holds for the closely related problem of computing a minimum
convex partition of a polygon, i.e., a covering with convex polygons
with pairwise disjoint interior. 
A variant of it was considered in the very first
Symposium on Computational Geometry~\cite{LubiwSoCG85}.
At the time, studying covering problems was motivated by applications 
in shape analysis and pattern recognition, graphics, and VLSI design.
Much of the focus was on the intrinsic 
complexity of the problem, less so on designing algorithmic solutions. 

O'Rourke was the first to show that the decision version of the problem is
indeed decidable~\cite{ORourkeDecidable}.
He shows how to construct an existential formula over
the reals which is true if and only if the given polygon has a convex cover
with $k$ pieces.
While polynomial-time algorithms have been developed for the 
minimum convex partition problem for simple polygons without 
holes~\cite{chazelle1985optimal},
O'Rourke and Supowit~\cite{ORourkeS83} proved the minimum convex cover problem 
to be \NP-hard for polygons with holes. 
Several years later, Culberson and Reckhow~\cite{CulbersonR94} showed that 
minimum convex cover by rectangles remains \NP-hard, even for
simple orthogonal polygons without holes.
Only recently, Abrahamsen~\cite{Abrahamsen21} 
managed to prove $\exists\mathbb{R}$-completeness of the convex cover problem (even when covering a simple polygon
by triangles), implying a negative answer
to the long-standing open problem of membership in \NP, unless $\NP=\exists\mathbb{R}$. 
While O'Rourke~\cite{ORourkeMCC82} already conjectured that optimal solutions 
may require irrational coordinates, 
Abrahamsen~\cite{Abrahamsen21} finally shows that Steiner points with
irrational coordinates of arbitrarily high algebraic degree can be necessary
for the corners of the pieces in an optimal solution for polygons with integral coordinates.
This makes it unlikely that exact methods (such as Integer Programming or Constraint
Satisfaction) can be employed in a straightforward manner.

In the context of the Challenge, only convex polygonal pieces with rational coordinates 
were allowed, both in order to avoid agreeing on a representation format for
algebraic numbers and in order to ease verification of submitted results.
Even in this constrained version, Steiner points with rational coordinates were still allowed. 
(We are not aware of any complexity
results that take this restriction into account for the general case 
of simple polygons with or without holes.)
Because the problem for general polygons was known to be \NP-hard, special
cases have been considered, most prominently orthogonal polygons 
to be covered with rectangles, as in the version that gave rise to the SoCG logo, Figure~\ref{fig:logo}.
However, the covering problem remains \NP-hard,
even for orthogonal polygons with holes~\cite{CulbersonR94};
in the pursuit of polynomially solvable versions of the problem, 
further special types of orthogonal polygons have been studied~\cite{Keil00}.

Besides restricting the allowed types of covering shapes from arbitrary
convex polygons to rectangles and axis-parallel rectangles, one can
also consider triangles. However, Christ~\cite{Christ11} has shown that this
version of the problem is also \NP-hard, 
and Abrahamsen~\cite{Abrahamsen21} shows it
to be $\exists\mathbb{R}$-complete as well.

Instead of convex polygons, other non-convex types of polygons 
can be considered for covering. 
If we consider star-shaped polygons, we are dealing with the 
Art Gallery Problem~\cite{ORourkeArtGallery}, 
another $\exists\mathbb{R}$-complete problem~\cite{AbrahamsenAM22},
for which optimal solutions may involve complicated algebraic coordinates.

\subsubsection{Simple Heuristics}
A first and simple heuristic approach is not to use Steiner points at all. 
Then the vertices of all convex pieces must also be vertices of the polygon 
to be covered. 
Furthermore, it suffices to consider only convex pieces $C$ that are maximal
in the sense that we cannot add another polygon vertex $v$, such that 
the convex hull of $C\cup \{v\}$ is larger than $C$, but still contained in $P$.
Therefore, a straightforward approach is to construct all (or some sufficient subset of) maximal convex polygons
formed by the set of vertices of $P$, and then use some search heuristic to select 
a subset covering $P$.
For a more general approach, Steiner points of a certain type can be added.
O'Rourke~\cite{ORourkeMCC82} suggests using the endpoints of maximal 
extensions of the 
edges of $P$ within $P$.
Continuing further, one may then add intersection points of these extension 
segments at a next stage, or, more generally, intersection points of the 
lines through pairs of Steiner points considered previously.  

\subsubsection{Approximate Solutions}
Another way to deal with a hard problem is to look at approximations.
Eidenbenz and Widmayer~\cite{EidenbenzW03} show the minimum convex 
cover problem to be APX-hard and provide an approximation algorithm 
with logarithmic approximation ratio. 
Their algorithm uses discretization and dynamic programming. 

There are different kinds of approximate solutions. Instead of approximating
the minimum number of convex polygons required to cover a polygon completely
one can relax the covering requirement as well and search for convex polygons
that cover the polygon approximately, e.g., by allowing a certain percentage 
of the area to stay uncovered or by considering convex polygons that overlap 
the exterior to some small extent. In robotics applications~\cite{AmiceDWZT22}, 
connectivity-preserving approximate covering with convex pieces is often 
sufficient. 

%

\subsection{Instances}
An important part of any challenge is the creation of suitable instances.
If the instances are easy to solve to optimality, the challenge becomes trivial;
on the other hand, if instances require a huge amount of computation for important common pre-processing steps
or for finding any decent solutions, the challenge may heavily favor teams that can afford better computation equipment.
The same is true if the set of instances becomes too large to manage with a single (or few) computers.

\begin{figure}
	\centering
	\begin{subfigure}[b]{.45\textwidth}
		\centering
		\includegraphics[width=\textwidth]{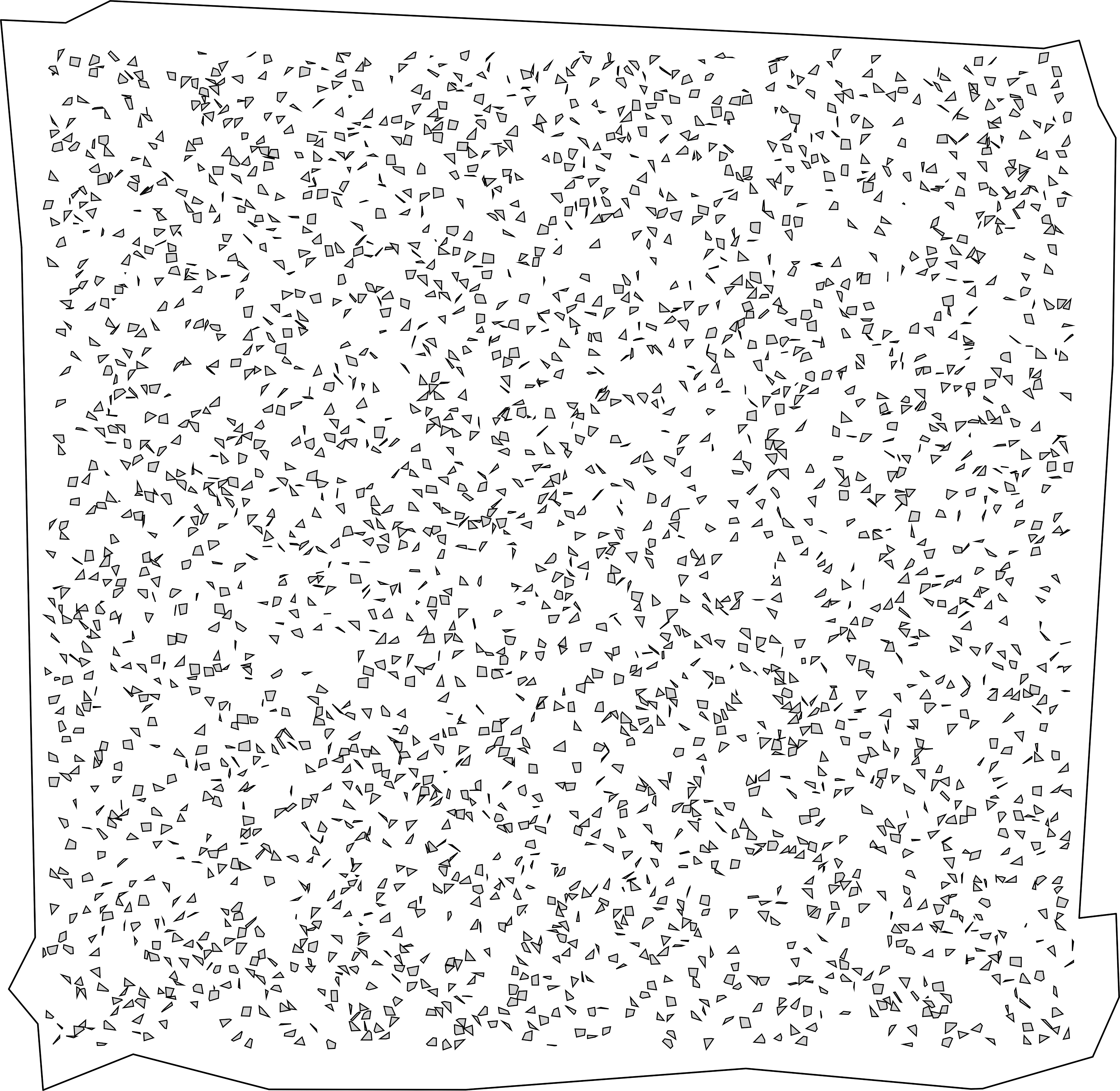}
		\caption{Instance \texttt{ccheese11045}.}
	\end{subfigure}\hfill
	\begin{subfigure}[b]{.45\textwidth}
		\centering
		\includegraphics[width=\textwidth]{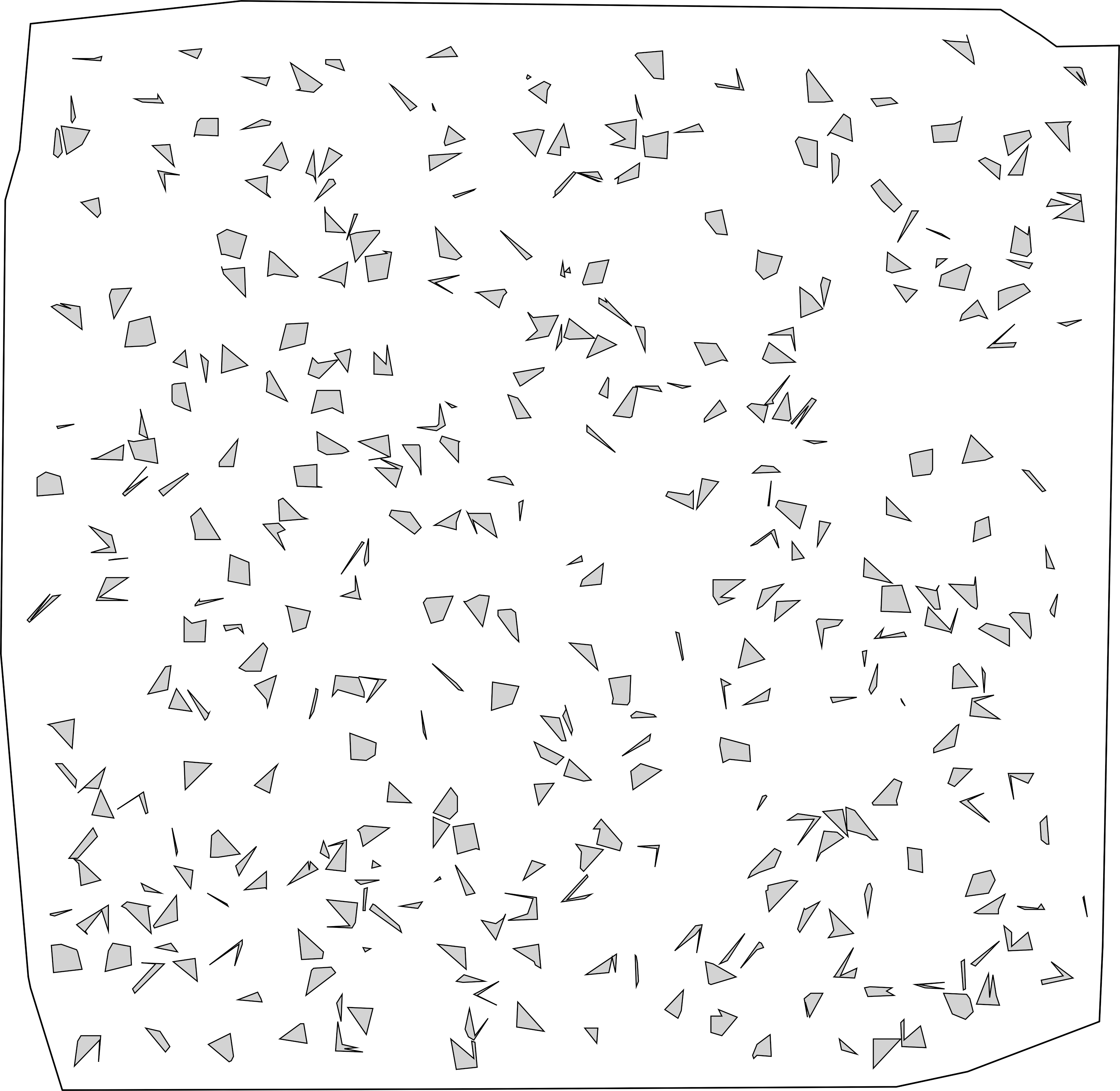}
		\caption{Instance \texttt{cheese1516}.}
	\end{subfigure}\hfill
	\begin{subfigure}[b]{.45\textwidth}
		\centering
		\includegraphics[width=\textwidth]{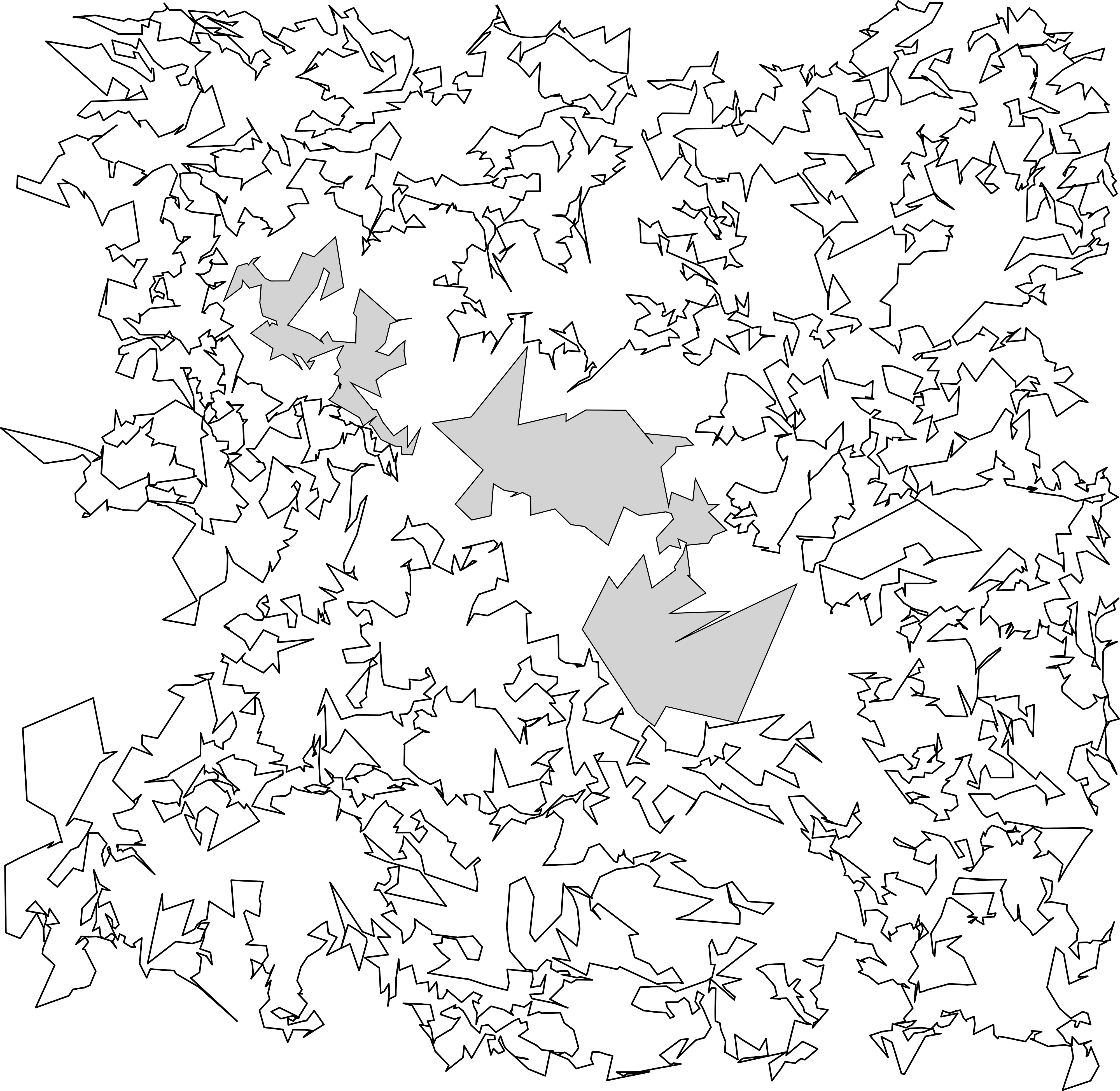}
		\caption{\texttt{fpg-poly\_0000004900\_h2}.}
	\end{subfigure}\hfill
	\begin{subfigure}[b]{.45\textwidth}
		\centering
		\vspace{0.3cm}
		\includegraphics[width=\textwidth]{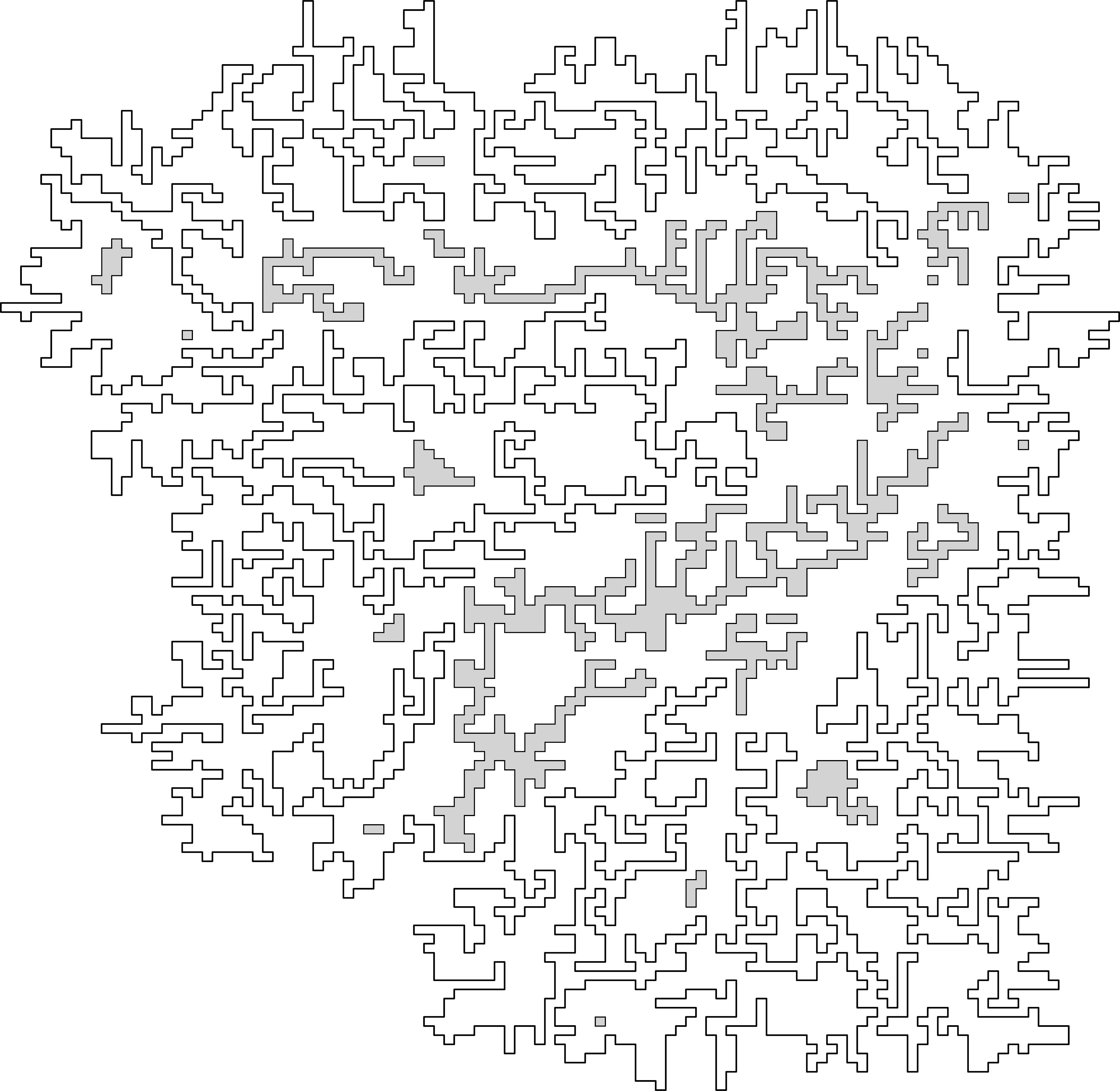}
		\caption{\texttt{srpg_iso_aligned_mc03255}.}
	\end{subfigure}\hfill
	\begin{subfigure}[b]{.45\textwidth}
		\centering
		\vspace{0.3cm}
		\includegraphics[width=\textwidth]{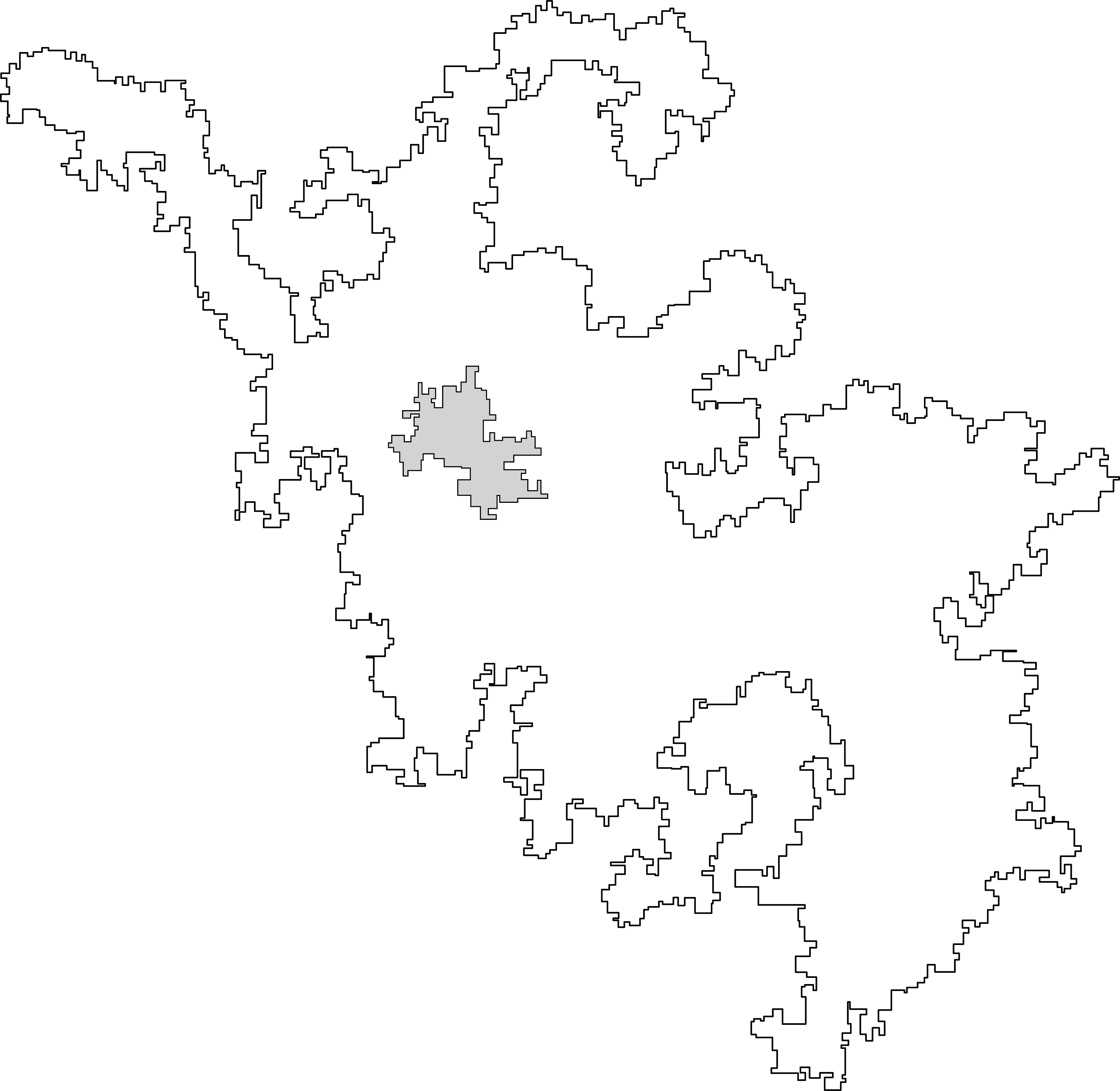}
		\caption{\texttt{srpg_iso_mc0001964}.}
	\end{subfigure}\hfill
	\begin{subfigure}[b]{.45\textwidth}
		\centering
		\vspace{0.3cm}
		\includegraphics[width=\textwidth]{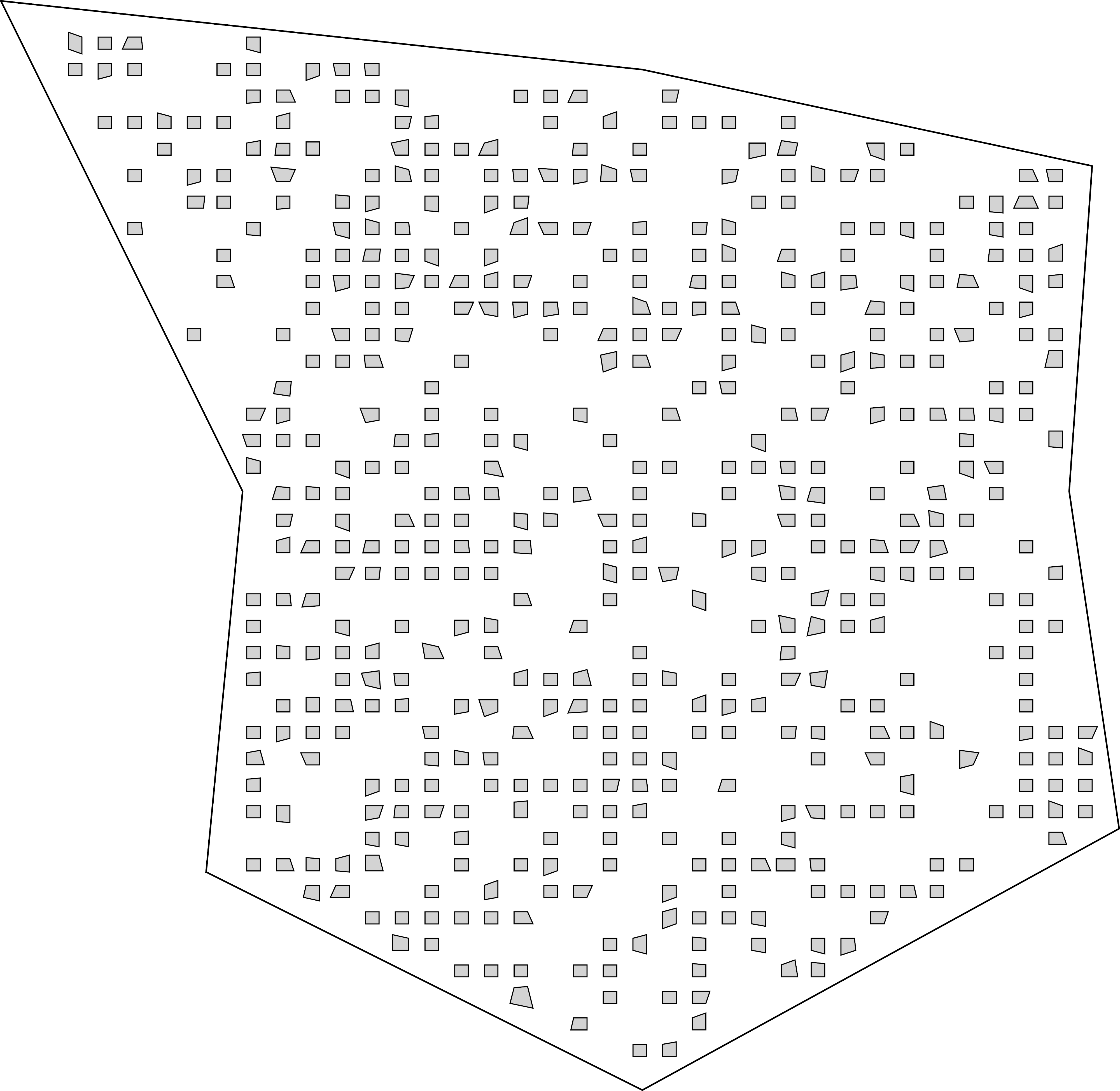}
		\caption{\texttt{maze_2534_250_05_01}.}
	\end{subfigure}
	\caption{A selection of actual contest instances made by different generators; gray areas are holes.}
	\label{fig:instance-examples}
\end{figure}

We used the following generators to generate our instances;
Figure~\ref{fig:instance-examples} shows corresponding examples of actual contest instances.

\begin{description}
	\item[cheese]
		The cheese generator's goal is to create a relatively simple outer boundary
		containing a large number of small holes. 
		To this end, we start by generating hole center points uniformly at random
		from a large rectangular region.
		We then choose a number of points on the boundary uniformly at random from
		a small range of possible values, usually $3$--$6$ points; all these points
		are chosen uniformly at random in the close vicinity of the hole center.
		They are then turned into a tour by visiting the points in an initially random order,
		which is turned into a polygon without self-intersections by applying a 2-opt step
		to reduce the tour length as long as there are intersecting edges.
		To make sure holes do not intersect or lie within other holes, each new hole
		is inserted into a 2D arrangement; if a hole intersects a previously added one,
		it is ignored and a new hole is generated around a new starting point instead.
		After the desired number of holes is generated, the outer boundary is generated
		by taking the convex hull of all hole center points and shifting it outwards where
		intersections with the holes make it necessary.
		Note that this may make the outer boundary non-convex.
	\item[ccheese]
		The ccheese generator works like the cheese generator, but with the additional goal of
		producing only convex holes;
		this is done by replacing each generated hole by the convex hull of its points.
		The reason behind this modification is the following.
		In particular for larger holes, non-convex vertices may often require a convex piece of their own
		to cover them.
		In many cases, such convex pieces can be turned into a maximal convex subregion
		of the feasible region in a unique way; this may inadvertently make large parts
		of the instance easy to solve.
	\item[srpg]
		Several instances are imported from the Salzburg Database of Polygonal Data~\cite{salzburgpoly}.
		The instances are taken from the database and normalized, so that all coordinates are non-negative.
		If the coordinates are not all integral, we obtain a similar polygon with integer coordinates
		by scaling all coordinates with a large factor and rounding them to the nearest integer.

		The \texttt{srpg}-family of instances is generated using their \emph{super random polygon generator}.
		Instances with prefix \texttt{srpg\_iso} are orthogonal polygons;
		the prefix \texttt{srpg\_iso\_aligned} indicates orthogonal polygons with integer coordinates
		which include many points with the same $x$- or $y$-coordinates.
		The prefix \texttt{srpg\_iso\_octa} indicates octagonal polygons, i.e., polygons where all angles are
		multiples of 45\degree.
		Finally, the prefixes \texttt{srpg\_smo} and \texttt{srpg\_smr} indicate random polygons
		with smoothed corners, i.e., polygons for which additional vertices are added to make
		corners smoother; \texttt{smr} indicates stronger smoothing than \texttt{smo}.
	\item[fpg]
		Like the \texttt{srpg} family, the \texttt{fpg} family of instances is taken from the
		Salzburg Database of Polygonal Data~\cite{salzburgpoly}, using the same approach
		to normalize and integralize the coordinates.
		These instances are generated using the FPG (triangulation perturbation) generator,
		which mutates an initial polygon, e.g., a regular polygon, by shifting its vertices
		while maintaining the boundaries' number of connected components.
		This often results in polygons with skinnier parts than what is usual in the \texttt{srpg} family.
	\item[maze]
		The maze generator generates polygons with a relatively simple outer boundary, into which
                        a large number of square obstacles are placed in a grid-like fashion,
			leaving long corridors of free space.
      These corridors require the use of highly overlapping polyongs to achieve good solutions.
      Some of the obstacles are then removed; others are
			slightly perturbed by moving some vertices of the obstacles outwards.
			These perturbations are meant to remove trivial approaches which cover each corridor with
			a single convex piece.
\end{description}

\subsection{Evaluation}
The contest was run on a total of \num{206} instances.

For many optimization problems, it is often considerably harder to find
a solution with the optimal value than it is to find a solution
that comes close to the optimum, say with value $\textrm{OPT}+c$ for some small constant $c$.

We suspected that this is the case for the contest problem as well.
In order to reflect this in the scoring of solutions,
instead of a score that linearly depends on the number of polygons, 
we introduced a quadratic scoring function.
For an instance $I$, let $B(I)$ be the number of convex pieces 
required in the best solution submitted for that instance.
Furthermore, let $T(I)$ be the number of convex pieces in the best solution of some team $T$ for $I$.
The score $S_T(I)$ of team $T$ for instance $I$ is
\[ S_T(I) := \frac{{B(I)}^2}{{T(I)}^2}. \]
As a consequence, doubling the number of convex pieces compared to the best known solution cuts the score down to $0.25$, and
all teams that submitted a solution for some instance $I$ that was not beaten by any other team receive a score of $1$ for $I$.
Teams that did not submit any valid solution for some instance $I$ receive a default score of 0 for $I$,
corresponding to a solution with an infinite number of convex pieces.
The total score $S_T = \sum_I S_T(I)$ of each team $T$ was then calculated by summing the scores of $T$ over all instances $I$.
The winner of the contest was the team with the highest score.
In case of ties, the tiebreaker was set to be the time a specific total score was obtained. 
As in previous years, this turned out to be unnecessary.

\subsection{Categories}
The contest was run in an \emph{Open Class}, in which participants could use any
computing device, any amount of computing time (within the duration of the
contest) and any team composition. In the \emph{Junior Class}, a
 team was required to consist exclusively of participants who were eligible
according to the rules of CG:YRF (the \emph{Young Researchers Forum} of CG Week), 
defined as not having defended a formal doctorate before 2021.

\subsection{Server and Timeline}
The contest itself was run through a dedicated server at TU Braunschweig,
hosted at \url{https://cgshop.ibr.cs.tu-bs.de/competition/cg-shop-2023/}.
It opened at 00:00 (UTC) on September 30, 2022 with a number of test instances,
with the full suite of contest instances released on October 28, 2022  and closed
at 24:00 (midnight, AoE), on January 27, 2023. 

During the contest, the code used by the server to verify submissions
was also made available to the participants as a python package on the Python Package Index (PyPI)\footnote{See \url{https://pypi.org/project/cgshop2023-pyutils/}.}.
Aside from trivial validity checks regarding encoding errors, the submissions were
also rigorously verified to be valid solutions to their respective instances.
This includes checking the convexity of all regions in each solution, 
checking that their union covers the entire area, and that no extra area is covered.
Using the CGAL~\cite{cgal} library, these checks are relatively straightforward to implement;
however, care must be taken when converting the floating-point, integer or rational numbers
submitted as solutions into CGAL's exact number types.
The massive amount of necessary Boolean operations on various polygons turned
out to be a serious stress test and actually helped unveil a
bug\footnote{\url{https://github.com/CGAL/cgal/issues/7235}} in CGAL's new
\texttt{join}-algorithm based on polylines that has since been addressed.

\section{Outcomes}
A total of 22 teams signed up for the competition, and 18 teams submitted
at least one valid solution.
In the end, the leaderboard for the top 10 teams looked as shown in Table~\ref{tab:top10}.
There were two teams (DIKU (AMW) and Shadoks) that were far ahead of all other participants.

\begin{table}[h!]
	\centering
	\begin{tabular}{|r|l|r|r|}
		\hline
		\textbf{Rank} & \textbf{Team} & \textbf{Score} & \textbf{Junior} \\
		\hline
		1 & DIKU (AMW) & 201.571 & \\
		2 & Shadoks& 198.347 & \\
		3 & BX23 & 144.150 & \\
		4 & SmartLab & 142.925 & \checkmark\\
		5 & agr & 122.400 & \\
		6 & rkPlayground & 121.311 & \checkmark\\
		7 & Karteflan & 103.392 & \checkmark \\
		8 & Ofir & 103.223& \\
		9 & pjgblt & 103.127 & \checkmark\\
		10 & cgI@tau & 100.893 &\\
		\hline
	\end{tabular}
	\caption{The top 10 of the final score, rounded to three decimal places.
	         Teams that satisfy the criteria for being considered a junior team
			 have a checkmark in the ``Junior'' column.}
	\label{tab:top10}
\end{table}

\begin{figure}
\centering
  \includegraphics[width=\textwidth]{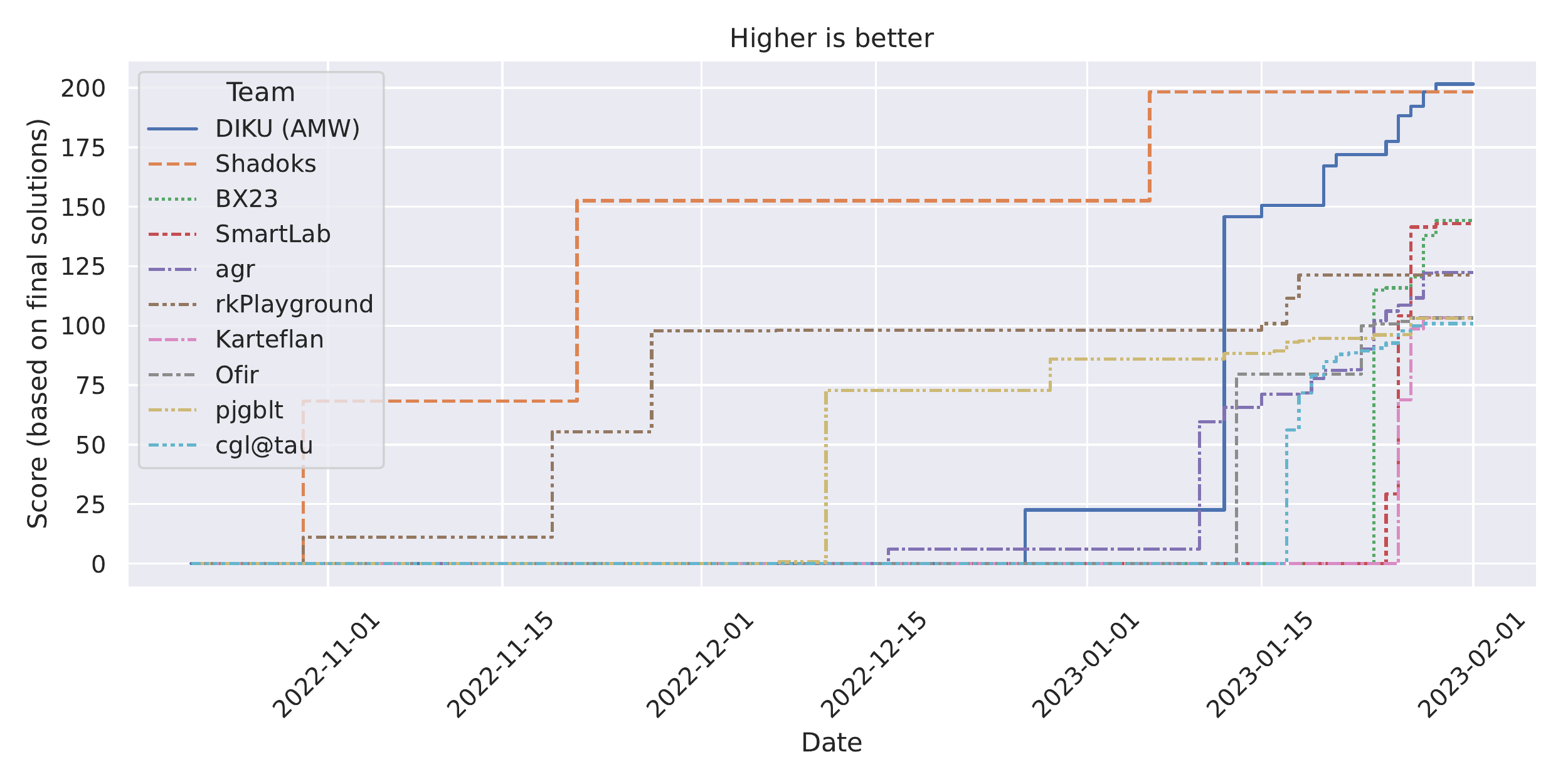}
  \caption{Score progress of the top 10 teams over time. The scores are computed based on the final submissions. Team \emph{Shadoks} maintained the lead until it was surpassed in the last days by team \emph{DIKU (AMW)} by a small margin. Most teams only started submitting serious submissions in the last two weeks. For many teams iterative improvements are visible.}
  \label{fig:score-progress}
\end{figure}

\begin{figure}
  \centering
  \includegraphics[width=\textwidth]{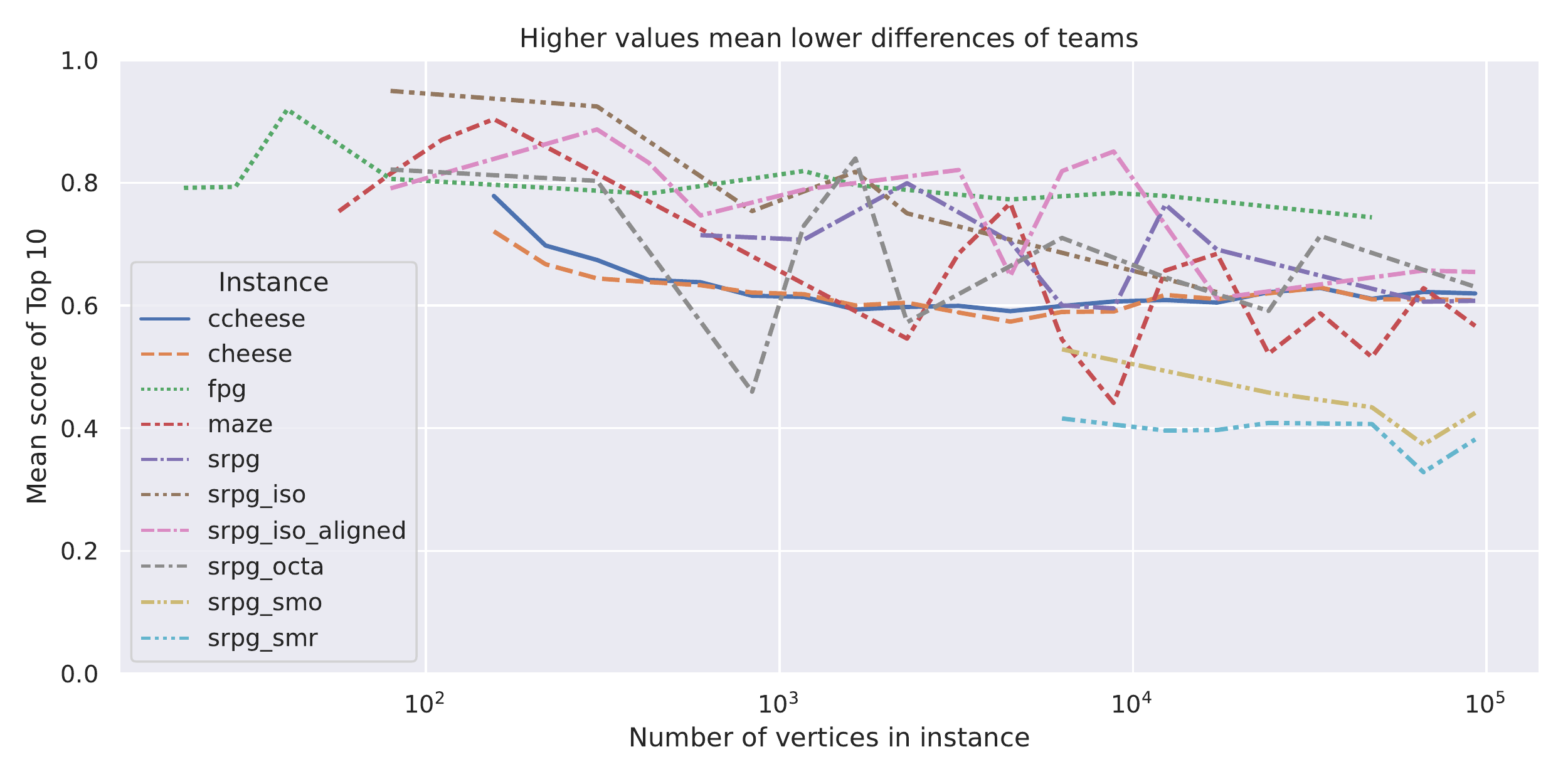}
  \caption{The mean score of the top 10 teams for the different instances gives an indication for their difficulty. The higher the mean score, the better the teams could keep up with the top teams. If the mean score is low this indicates that additional, non-trivial ideas were necessary to keep up. The first visible challenge the teams likely had is scalability, as the mean score decrease for larger instances.  Additionally, the \texttt{ccheese} and \texttt{cheese} instances seem to have been especially challenging even at medium size, while being of similiar difficulty themselves. For larger instances, the \texttt{srpg} and \texttt{maze} instances show to be challenging, with the smoothed \texttt{srpg}-instances \texttt{smo} and \texttt{smr} being the most challenging instances of all. The \texttt{fpg} instances seem to have been relatively easy.}
  \label{fig:scores_over_size}
\end{figure}

The progress over time of each team's score can be seen in Figure~\ref{fig:score-progress};
the best solutions for all instances (displayed by score) can be seen in Figure~\ref{fig:scores_over_size}.
The top two teams were invited for contributions in the 2023 SoCG proceedings, as follows.

\begin{figure}
  \centering
  \includegraphics[width=\textwidth]{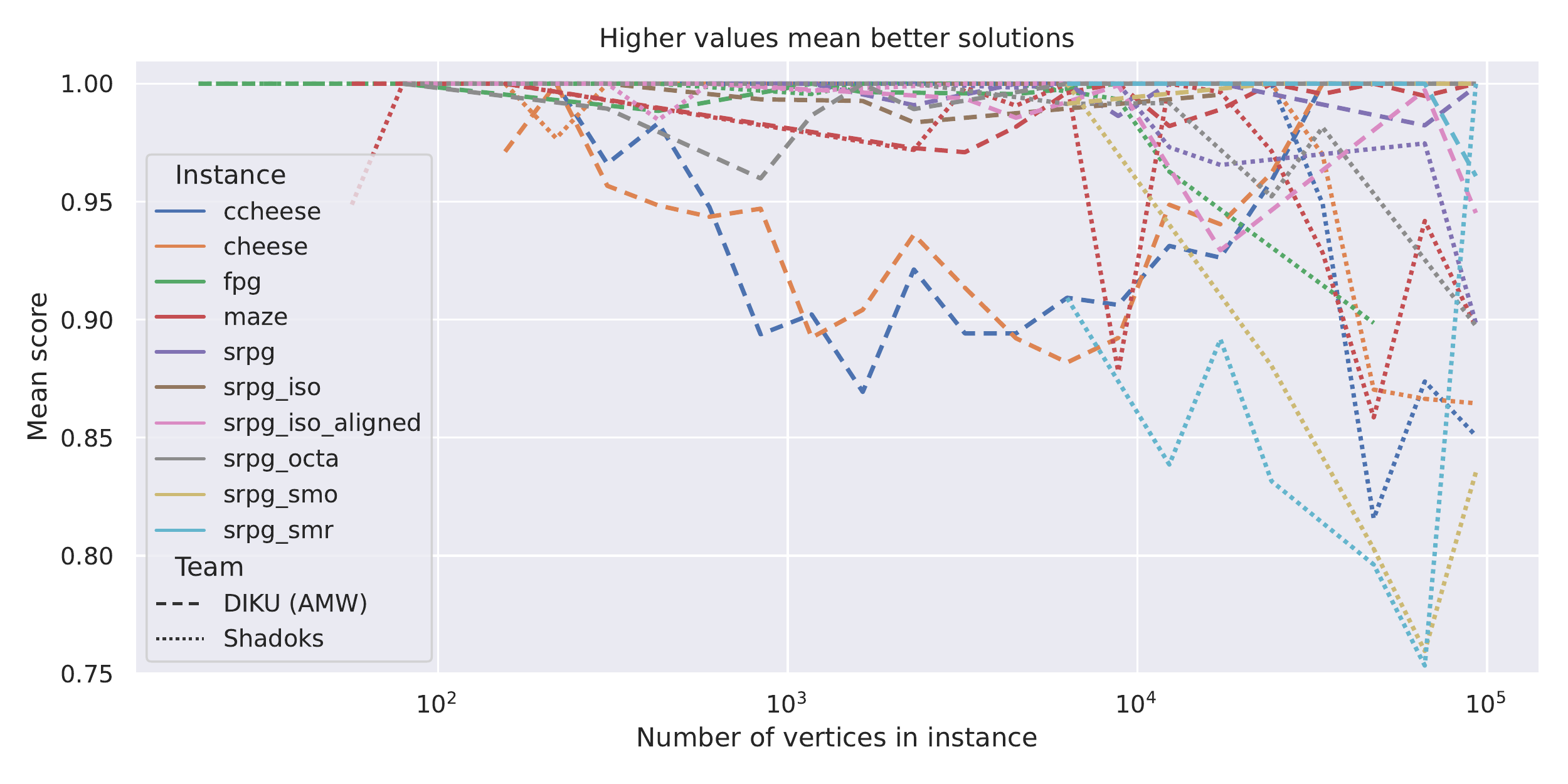}
  \caption{The two top teams have a relatively similar score, but they show different strengths. While team \emph{Shadoks} performs better for the \texttt{cheese} and \texttt{ccheese} instances, the approach of \emph{DIKU (AMW)} seems to scale better, such that they take the lead for large instances. For the \texttt{maze} instances a varying lead for different instances is visible.}
  \label{fig:two_top}
\end{figure}

\begin{enumerate}
\item Team DIKU (AMW): Mikkel Abrahamsen, William Bille Meyling, Andr\'e Nusser~\cite{diku}.
\item Team Shadoks: Guilherme D. da Fonseca~\cite{shadoks}.
\end{enumerate}
Details of their methods and the engineering decisions they made are given in their respective papers.
Their strengths and weaknesses are shortly evaluated in Figure~\ref{fig:two_top}.
In the following, we give a very brief description of their approaches.

Team DIKU (AMW)~\cite{diku} bases their approach on a constrained Delaunay triangulation of the
vertices of the given polygon $P$ along with some additional points, e.g., intersections
of extensions of the segments of $P$.
On this triangulation, they compute a \emph{visibility graph}, which has
a vertex for each triangle and an edge between pairs of triangles whose convex hull
is completely contained in $P$.
They observe that cliques in this graph correspond to convex polygons induced 
by the triangles in the clique.
Sometimes, these induced polygons need not be fully contained in $P$, but they
assume that such situations are relatively rare.
They then use an existing implementation called ReduVCC~\cite{reduvcc} due to Strash and Thompson
for the vertex clique cover problem to compute a small number of cliques that cover all triangles.
Finally, they repair any convex pieces that are not fully contained in $P$ and 
remove any pieces that only cover parts of $P$ that are already covered by other pieces.

Team Shadoks~\cite{shadoks} had a different approach.
On a high level, their approach consists of generating a so-called \emph{collection} $C$,
which is a set of convex pieces that are contained in $P$ and together cover all of $P$.
While it is important to not let $C$ grow too large, the goal for generating a good $C$
is only to have a small \emph{solution} $S \subseteq C$ which also covers all of $P$.
Given a good $C$, Shadoks transform the task of finding a good solution $S \subseteq C$
into a moderately sized instance of the Set Cover problem, which is either solved
optimally using an integer programming solver or heuristically using simulated annealing;
in order to do this, they introduce witness points in $P$ and enforce that each witness
be covered by at least one set in $S$.
They propose several methods to generate $C$; one is based on a modified version of the
Bron-Kerbosch algorithm~\cite{bronkerbosch} to enumerate maximal cliques,
the other is a procedure they call \emph{random bloating}.

In order to evaluate what score a relatively simple, straightforward approach to the problem
would achieve, we implemented the following type of heuristic.
Starting from a constrained Delaunay triangulation of the vertices of the polygon $P$,
we greedily merge arbitrarily chosen faces of the current subdivision
to form convex pieces with which to cover $P$, until no more faces can be merged.
Our implementation of this scheme would have achieved a score of \num{85.3} and 
thus would have ranked 12th in the challenge.

This shows the massive advantage of the approaches of the top teams over simple, straightforward methods;
it also shows that the majority of actual contest participants actually came up with algorithms that 
were able to beat such methods.

\section{Conclusions}

The 2023 CG:SHOP Challenge motivated a considerable number of teams to engage in extensive optimization studies.
The outcomes promise further insight into the underlying, important optimization problem.
Moreover, the considerable participation of junior teams indicates that the Challenge itself
motivates a great number of students and young researchers to work on practical algorithmic problems.

\bibliography{bibliography}
\end{document}